\documentclass[twocolumn, pra, showpacs,superscriptaddress]{revtex4}%
\usepackage{graphicx}
\usepackage{amsmath}
\usepackage{bm}
\usepackage{float}
\usepackage{color}
\usepackage{sidecap}
\usepackage{soul}
\usepackage[colorlinks, citecolor=blue]{hyperref}
\usepackage{amsfonts}
\usepackage{amssymb}%
\providecommand{\U}[1]{\protect\rule{.1in}{.1in}}
\setstcolor{red}
\begin{document}
\title{Theoretical Simulation of $^{87}$Rb Absorption Spectrum in a Thermal Cell}
\author{Cheng Hong}
\affiliation{State Key Laboratory of Magnetic Resonance and Atomic and Molecular Physics,
Wuhan Institute of Physics and Mathematics, Chinese Academy of Sciences, Wuhan
430071, People's Republic of China}
\affiliation{University of Chinese Academy of Sciences, Beijing 100049, People's Republic
of China}
\author{Zhang Shan-Shan}
\affiliation{State Key Laboratory of Magnetic Resonance and Atomic and Molecular Physics,
Wuhan Institute of Physics and Mathematics, Chinese Academy of Sciences, Wuhan
430071, People's Republic of China}
\affiliation{University of Chinese Academy of Sciences, Beijing 100049, People's Republic
of China}
\author{Xin Pei-Pei}
\affiliation{State Key Laboratory of Magnetic Resonance and Atomic and Molecular Physics,
Wuhan Institute of Physics and Mathematics, Chinese Academy of Sciences, Wuhan
430071, People's Republic of China}
\affiliation{University of Chinese Academy of Sciences, Beijing 100049, People's Republic
of China}
\author{Cheng Yuan}
\affiliation{State Key Laboratory of Magnetic Resonance and Atomic and Molecular Physics,
Wuhan Institute of Physics and Mathematics, Chinese Academy of Sciences, Wuhan
430071, People's Republic of China}
\affiliation{University of Chinese Academy of Sciences, Beijing 100049, People's Republic
of China}
\author{Liu Hong-Ping\thanks{Email:liuhongping@wipm.ac.cn}}
\affiliation{State Key Laboratory of Magnetic Resonance and Atomic and Molecular Physics,
Wuhan Institute of Physics and Mathematics, Chinese Academy of Sciences, Wuhan
430071, People's Republic of China}
\affiliation{University of Chinese Academy of Sciences, Beijing 100049, People's Republic
of China}

\begin{abstract}
In this paper, we present a theoretical simulation of $^{87}$Rb absorption
spectrum in a thermal cm-cell which is adaptive to the experimental
observation. In experiment, the coupling and probe beams are configured to
copropagate but perpendicular polarized, making up to five velocity selective
optical pumping (VSOP) absorption dips able to be identified. A $\Lambda$-type
electromagnetically induced transparency (EIT) is also observed for each group
of velocity-selected atoms. The spectrum by only sweeping the probe beam can
be decomposed into a combination of Doppler-broadened background and three
VSOP dips for each group of velocity-selected atoms, companied by an EIT peak.
This proposed theoretical model can be used to simulate the spectrum adaptive
to the experimental observation by non-linear least-square fit method. The fit
for high quality of experimental observation can determine valuable transition
parameters such as decaying rates and coupling beam power accurately.

\end{abstract}
\keywords{Electromagnetically induced transparency, velocity selective resonance,
optical pumping, Multilevel system}
\pacs{42.50.Gy, 32.70.Jz, 32.10.Fn, 32.80.Xx}
\date{\today }
\maketitle
\volumeyear{ }
\volumenumber{ }
\issuenumber{ }
\eid{ }
\received[Received text]{}

\revised[Revised text]{}

\accepted[Accepted text]{}

\published[Published text]{}

\startpage{1}
\endpage{ }

\section{Introduction}

Laser-atom interaction can control the quantum interference properties between
atomic states, which has various significant applications in many fields. For
example, recently, coherence and interference effects in atomic systems such
as coherent population trapping (CPT) \cite{968,969}, refractive index
enhancement \cite{973}, electromagnetically induced transparency (EIT)
\cite{688, 450, 1012, 963, 899, 877, 1193, 1201, 1202}, electromagnetically
induced absorption (EIA) \cite{965, 466, 537, 692, 531, 549, 519} and velocity
selective optical pumping (VSOP) \cite{891, 981, 980} have been widely applied
in atomic clock \cite{974}, squeezing of light \cite{977}, light storage
\cite{976, 975} and quantum computation \cite{978}. Particularly, the theoretical
and experimental investigations on EIT in atomic systems are of enhanced
interest in scientific researches. EIT was also studied in the atomic
ensemble\cite{1202}, atom-molecule systems \cite{1201}, and solid-state
systems \cite{1215, 1210, 1208, 1204}.

Unlike EIT in simple three-level $\Lambda$, $V$ and cascade (ladder)-type
atomic systems \cite{692, 724, 912, 1212, 1213}, however, most alkali atoms
have a complicated energy level structure instead of following an ideal model
and the Doppler broadening causes many states involved in the transitions, for
example, four \cite{518,525}, five \cite{983, 984}, even six level \cite{979,
546} systems having been studied. The $D_{2}$ transition of Rb consisting of
two ground hyperfine levels and four excited levels forms a six-level scheme
system. The separations of the upper hyperfine levels are less than the
Doppler broadening of the transition in the room temperature, causing more
extra satellite dips also observed due to Doppler shifted along with the EIT
peak in the probe transmission profile \cite{928,980,981}.

These velocity selective resonances result in complexity of the final spectrum
and difficulty in their analysis. For example, Maguire \textit{et al}.
numerically calculated the optical Bloch equations by taking into the optical
effects and the simulated spectra could reproduce the main features of the
observed saturation absorption spectra of $^{85}$Rb $D_{2}$ transition
\cite{985}. Bhattacharyya \textit{et al}. studied velocity selective resonance
dips along with EIT peak observed in the experiment by solving the density
matrix equations of a $\Lambda$-type five-level system \cite{498, 472, 609,
479,1012}. Applying the perturbation method to the optical Bloch equations,
valuable information for induction of EIA was obtained for a closed multilevel
$F_{g}=1\rightarrow F_{e}=2$ transition in the Hanle configuration \cite{928}.
Similarly, by solving the rate equations, Krmpot \textit{et al}. could well
identify spectral position and intensity for atoms with different velocities
\cite{942}. Ray \textit{et al}. carried out a detailed theoretical analysis of
the coherent process by solving the density matrix formulation including all
orders of pump and probe powers without any assumptions \cite{951}.

However, all of the theories and simulations have to depend on a series of accurate given
dynamical parameters such as decay rates and temperature, etc.. Due to the complexity of the atomic multi-levels
and uncertainty in laser parameters, most of time, it is difficult to present a simulation comparable with the experiment with high quality.
In this paper, rather than starting from the theory based on the Bloch
equations, we propose a semiempirical model to numerically explain the
experimental observation of the multi-level system of $^{87}$Rb D$_{2}$ line.
The model with varied dynamical parameters can be adaptive to the experimental observation and
 can give exact coincidence with the observed spectral profile and details.
The EIT peak signal stemming from the background of one VSOP absorption dip is
also well resolved. The theoretical model is constructed from the main
concerned physical processes in the $\Lambda$-type EIT with both coupling and
probe beams copropagating but perpendicularly polarized, namely, the spectral analysis should consider the Doppler-broadening
and Doppler-free processes at the same time.

\section{Experiment}

\begin{figure}[ptb]
\begin{center}
\includegraphics[width=8cm]{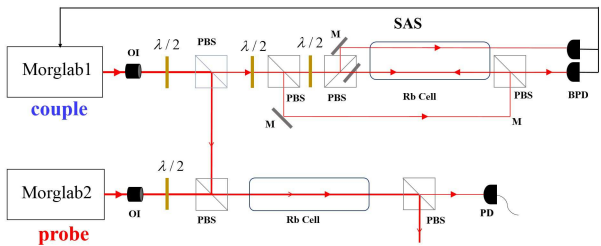}
\end{center}
\caption{(Color online) Experimental setup for EIT spectral measurement. PBS:
polarizing beam splitter; OI: optical isolator;~M: mirror; $\lambda/ 2$: half
wave plate; $\lambda/ 4$:~quarter wave plate;~PD: photo detector; SAS:
saturation absorption spectroscopy setup.}%
\label{fig1}%
\end{figure}

Many works have been done for the observation of the $\Lambda$-type EIT for
$^{87}$Rb \cite{473, 477, 425}, but to obtain high quality of spectral data
for the theoretical analysis, we have to re-investigate the experimental
observation. The experimental setup is shown in Fig. \ref{fig1}. Two
filter-tuned, cateye refletor feedback external cavity diode lasers (Moglabs)
with line-widths $<1$ MHz are used in the experiment. One is used to couple
hyperfine levels of the ground $^{5}S_{1/2}$ and excited $^{5}P_{3/2}$ states
of the $^{87}$Rb isotope ($D_{2}$ line in Rb), and the other one serves as a
prober to monitor the transparency of the laser through the Rb vapor. The pump
laser beam is split into two parts where the part transmitted through a
polarizing beam splitter (PBS) is sent to a Doppler-free saturation absorption
set-up (SAS) for locking the laser frequency. At the same time, the pump laser
is also partly reflected by the PBS and sent through the sample cell for the
EIT experiment as coupling beam. The pumping laser can be locked to any
possible hyperfine and crossover transition peak with a long term frequency
stability less than $1$ MHz.

In a similar way, the probe laser beam transmitted through the PBS serves as
the probe beam in the EIT experiment. The coupling and probe laser beams
copropagate through the Rb cell, and their polarizations are linear and
mutually perpendicular. The coupling and probe laser beams are adjusted to
overlap almost completely throughout the total length of the cell. The
coupling beam size is around $2.5$ mm$^{2}$. After the Rb cell, the probe beam
was extracted by another PBS and detected by a photodiode. The transmission of
the probe laser was detected with its frequency swept across the $^{2}P_{3/2}$
levels from the ground state $F=1$. The laser intensities can be controllable
by combination of half-wave plate and PBS.

The cylindrical shaped Rb vapor cell is made of pyrex glass and has a size of
length $7.5$ cm and diameter $2.5$ cm. The pressure inside is $10^{-6}$ Torr
at room temperature ($\sim25%
{{}^\circ}%
$C) without any buffer gas. The cell filled with both isotopes of Rb in their
natural abundances $^{85}$Rb ($72{\%}$) and $^{87}$Rb ($28{\%}$). We have not
applied any magnetic field shielding outside the Rb vapor cell since the
energy level shifting induced by the earth's magnetic field ($\sim0.5$ Gauss)
is less than $0.5$ MHz.

\section{Theory}

\begin{figure}[ptbh]
\begin{center}
\includegraphics[width=8cm]{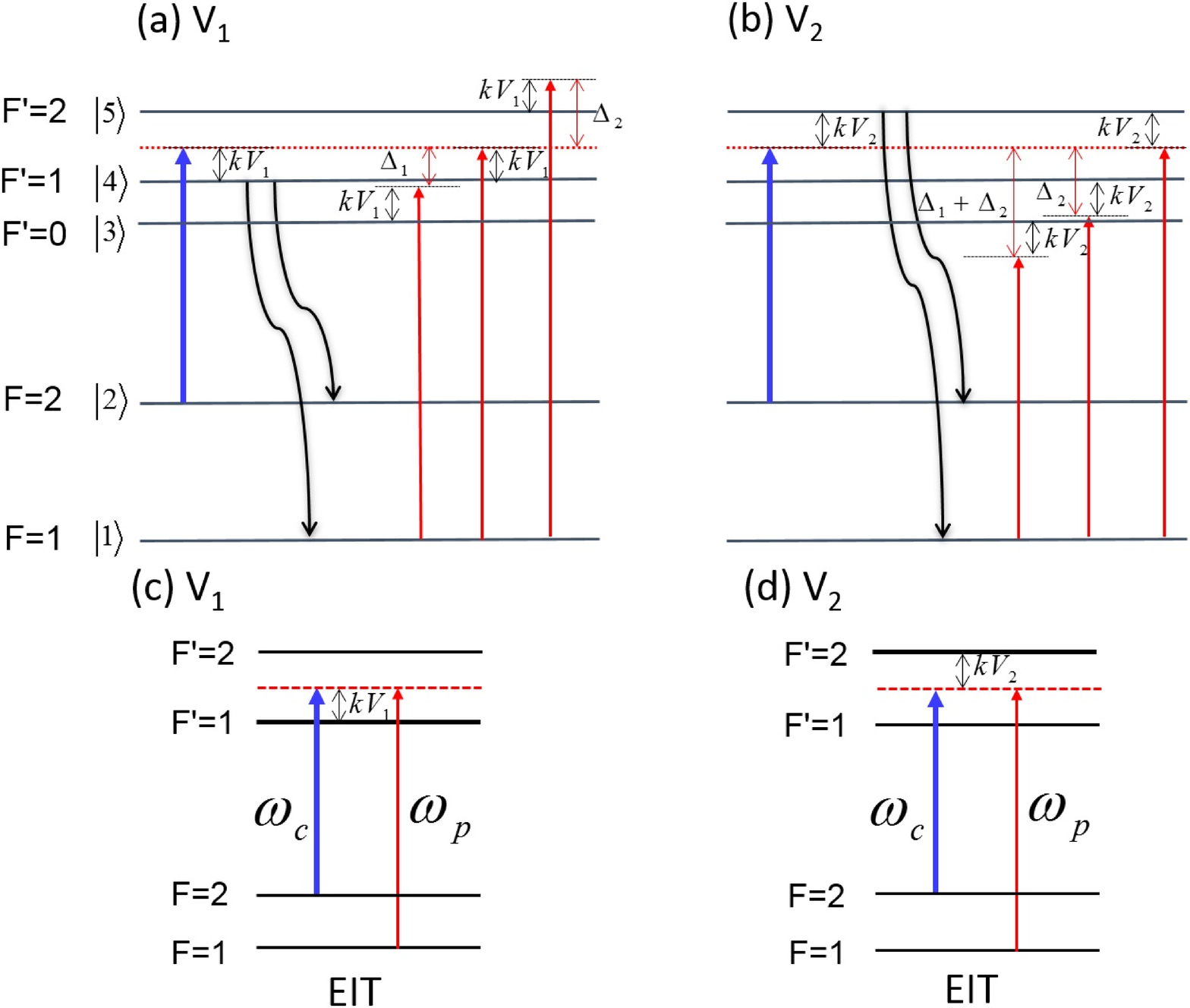}
\end{center}
\caption{(Color online) The energy diagram and spectral assignment of the
observed spectrum. (a) and (b) are the level schemes for the $V_{1}$ and
$V_{2}$ velocity groups, respectively. The blue solid line indicates the
couple beam frequency $\omega_{c}$ locked at the $F=2$ to $F^{\prime}%
=$CO$_{1-2}$ while the weak (red colored) line indicates the probe beam
scanning from the ground level $F=1$ to the excited level $F^{\prime}=0,1,2$.
(c) and (d) are the level schemes to show the formation of the EIT for $V_{1}$
and $V_{2}$ velocity groups. }%
\label{fig2}%
\end{figure}

We considered a $\Lambda$-type five-level atomic system interacting with two
lasers as shown in Fig. \ref{fig2}. The levels $\left\vert 1\right\rangle
,\left\vert 2\right\rangle $ are the two hyperfine levels in the ground state
$^{5}S_{1/2}$ and the levels $\left\vert 3\right\rangle ,\left\vert
4\right\rangle ,\left\vert 5\right\rangle $ are the three closely spaced
hyperfine levels in the excited state $^{5}P_{3/2}$. The strong coupling laser
is frequency locked to one of the ground level $\left\vert 2\right\rangle $ to
$\rightarrow\left\vert j\right\rangle $ ($j=4,5$) or their crossover
transition. While the weak probe beam scans over all the excited levels
$\left\vert 3\right\rangle ,\left\vert 4\right\rangle ,\left\vert
5\right\rangle $ from the other ground level $\left\vert 1\right\rangle $,
corresponding to transitions from $F=1$ to $F^{\prime}=0,1,2$ states. The
transitions $\left\vert 1\right\rangle \rightarrow\left\vert j\right\rangle $
($j=3,4,5$) and $\left\vert 2\right\rangle \rightarrow\left\vert
j\right\rangle $ ($j=4,5$) are dipole-allowed while the transitions between
the other levels are dipole-forbidden \cite{977}.

Due to Doppler broadening, the frequency-fixed coupling laser can populate two
distinct velocity groups of atoms to the upper hyperfine levels $F^{\prime
}=1,2$ from the ground level $F=2$, denoted as $V_{1}$ and $V_{2}$,
respectively. These populations on excited states will decay to the two ground
hyperfine levels $F=1,2$ by spontaneous emission. We are interested in the
velocity selective population on the $F=1$ ground level. The copropagated weak
probe laser has an additional response for these velocity selective atoms.
This additional response is superimposed on a Doppler background. Finally,
VSOP dips can be observed in the probe transmission profile at Doppler shifted
frequencies, as well as EIT signals.

Considering the VSOP signal is much narrower than the Gaussian-broadening, we
can omit the convoluting process as an approximation. So here we only consider
the convolution of EIT with Gaussian-broadening. Therefore, we can approximate
the final spectrum as the sums of absorption VSOPs and transparency EIT
superimposed on the Doppler broadening. The Doppler background is
mathematically expressed by Gaussian distribution function%
\begin{equation}
G(\omega_{p})\propto e^{-\frac{v^{2}}{%
\operatorname{u}%
^{2}}} \label{eq1}%
\end{equation}
with the most probable velocity defined to be $u=(2k_{B}T/m_{Rb})^{1/2}$ and
$v$ satisfying the relation $\omega_{p}=\omega_{1j}(1-\frac{v}{c})$.

The EIT signal has the form \cite{1004}%
\begin{align}
EIT(\Delta_{p}) &  \propto\operatorname{Im}\int\frac{1}{\frac{(\Omega
_{c}/2)^{2}}{(\Delta_{p}+\Delta_{c})+\frac{\omega_{p}-\omega_{c}}{c}%
v+i\gamma_{c}}-(\Delta_{p}+\frac{\omega_{p}}{c}v+i\gamma_{p})}\nonumber\\
&  \times e^{-\frac{v^{2}}{%
\operatorname{u}%
^{2}}}dv\label{eq2}%
\end{align}
with Doppler broadening considered. If the Doppler-broadening can be ignored, we can easily set
the Gaussian-broadening to zero and the formula is reduced to a standard Doppler-free EIT. Here $\Omega_{c}$\ is the Rabi frequency
for the coupling laser, $\Delta_{p}$ and $\Delta_{c}$ the detunings of the
probe and coupling laser fields, respectively. Taking the Doppler velocity
selective effect into account, $\Delta_{c}=0$ for the selected atoms and the
EIT occurs at $\omega_{p}=\omega_{c}+\Delta$, where $\Delta=6.835$ GHz is the
hyperfine level splitting of the ground state. The numerical skills are used
for the integral in Eq. \ref{eq2} \cite{1004}.

There are three VSOP dips corresponding to the transition $\left\vert
1\right\rangle \rightarrow\left\vert j\right\rangle $ ($j=3,4,5$) for each
velocity group of atoms, and the VSOP has the form of Lorentzian line-type
\cite{472, 477}%

\begin{equation}
L(\omega_{p})\propto\frac{1}{(\omega_{p}-\omega_{1j})^{2}+\Gamma_{p}^{2}},
\label{eq3}%
\end{equation}
where $\omega_{1j}$ ($j=3,4,5$) represents the energy gap corresponding to the
transition $\left\vert 1\right\rangle \rightarrow\left\vert j\right\rangle $
($j=3,4,5$).

The observed spectrum for every group of velocity can be viewed as the
superpositions of these three basic line types listed in Eqs. \ref{eq1}%
-\ref{eq3}
\begin{align}
Spectrum  &  =a_{1}\times EIT(\Delta_{p})\oplus\sum\limits_{j=3}^{5}%
a_{2j}\times L(\omega_{p})\nonumber\\
&  \oplus\sum\limits_{j=3}^{5}a_{3j}\times G(\omega_{p}),
\end{align}
where $a_{1}$, $a_{2j}$ and $a_{3j}$ are the combination coefficients.

This formula has a simpler form and it can be easily used to make an analysis
for the observed experimental data, especially to perform a least-square fit
for extracting some spectral character parameters.

\section{Results and discussion}

\begin{figure}[ptbh]
\begin{center}
\includegraphics[width=9.5cm] {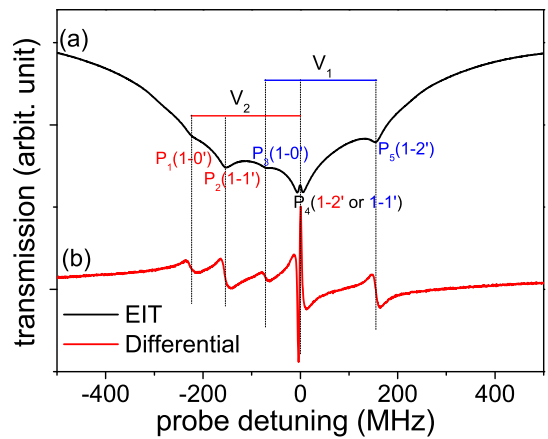}
\end{center}
\caption{(Color online) Experimental trace of the five VSOPs and the
corresponding differential signal in the $\Lambda$-type $^{87}$Rb system with
the coupling beam locked to $F^{\prime}= $CO$_{12}$ at room temperature, while
the weak probe laser scans over all the transitions ($F=1 \to F^{\prime
}=0,1,2$). }%
\label{fig3}%
\end{figure}

Figure \ref{fig3} presents the probe transmission spectrum as a function of
the probe detuning and the corresponding derivative signal for $^{87}$Rb
$D_{2}$ transition. The power of the pump and probe lasers are around $5$ and
$0.005$ mW, respectively. For the $^{87}$Rb-$D_{2}$ line, the separation of
the two hyperfine levels ($F=1,2$) in the ground state is $6.835$ GHz, while
the separations between the hyperfine levels ($F^{\prime}=0,1,2,3$) of the
excited level are $\Delta_{1}=72.3$ MHz (for $F^{\prime}=0$ and $1$),
$\Delta_{2}=157.2$ MHz (for $F^{\prime}=1$ and $2$) and $\Delta_{3}=267.1$ MHz
(for $F^{\prime}=2$ and $3$), respectively \cite{1173}. We observed five VSOP
dips (P$_{1}$,P$_{2},...$P$_{5}$) and a narrow EIT peak (P$_{4}$) superimposed
on a Doppler background. The separations of the three velocity-selective
absorption dips (P$_{1}$, P$_{2}$, P$_{3}$) on the left side of the EIT peak
(P$_{4}$) are $-(\Delta_{1}+\Delta_{2})$, $-\Delta_{2}$, and $-\Delta_{1}$
with respect to the reference frequency $\omega_{p}-\omega_{c}-\Delta=0$.
Another dip (P$_{5}$) on the right-hand side is at a distance of $\Delta_{2}$
from the EIT peak.

\begin{figure}[ptbh]
\begin{center}
\includegraphics[width=7cm] {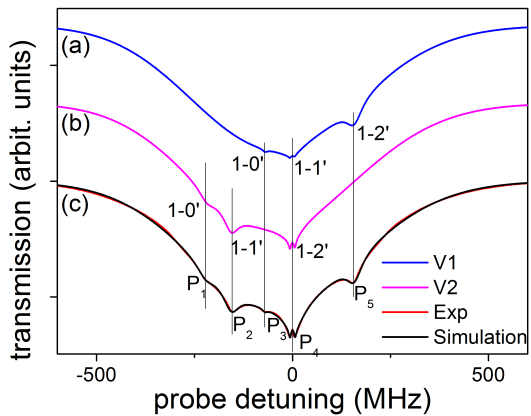}
\end{center}
\caption{(Color online) Spectral simulation for the experimental observation.
The spectrum can be decomposed into two velocity selective groups $V_{1}$ (a),
and $V_{2}$ (b). The simulation of the total probe absorption spectra is in
good agreement with the observation (c).}%
\label{fig4}%
\end{figure}

We can have an analysis for the observed spectrum, namely, the spectrum can be
decomposed into the overlapping of two groups of different velocity selection.
It is shown in Fig. \ref{fig4}. For each velocity group, the transmission
spectrum of the probe laser is modeled as a sum of absorptions of three VSOPs,
relatively weighted by a Maxwell-Boltzmann distribution, and one EIT over the
Doppler-broadening, as described in the theory part. In our process, all the
weights are replaced by the combination coefficients and assigned to some
trial values by visional comparison of the simulation and the experimental
trace. They are varied in the least-square fit and arrive at convergent values
by iteration algorithm. By switching on and off the corresponding
coefficients, we can plot the respective spectral simulations corresponding to
the two different velocity groups of atoms ($V_{1}$ and $V_{2}$), which are
shown in Fig. \ref{fig4}(a) and (b). They are summarized again to produce the
final simulated absorption spectrum as shown in Fig. \ref{fig4}(c), which
coincides with the experimental observation with high quality. From the
convergent fitting, we can determine the parameters of $\gamma_{c}$,
$\gamma_{p}$ for the EIT. The determined values are $\gamma_{c}=3$ MHz and
$\gamma_{p}=6$ MHz, in good agreement with $\gamma_{c}=(\gamma_{j2}%
+\gamma_{21})/2=3$ MHz, $\gamma_{p}=(\gamma_{j2}+\gamma_{j1})/2=6$ MHz
\cite{1132}, in which $\gamma_{j1}$ and $\gamma_{j2}$ equal to the natural
decay rate $\Gamma$ (6 MHz), $\gamma_{21}$ is the nonradiative decay rate
between the ground levels ($100$ kHz) which is negligible. The parameter
$\Omega_{c}$ can also be determined from the EIT peak. It has a value
$\Omega_{c}=14$ MHz, close to $\Omega_{c}=12.7$ MHz estimated from laser power
$5$ mW and beam diameter $d=0.2$ cm. In the estimation, the formula
$\Omega_{c}=\mu E/\hbar$ is used, but the beam has an irregular shape and it
is difficult to estimate the effective diameter. It should be noted that we
can also have fitted $\Gamma_{p}$ values for VSOP peaks in every velocity
group. Their values are $17$, $22$ and $25$ MHz, respectively. In the fit, the
parameters of line center $\omega_{1j}$ in Eq. \ref{eq1} are fixed and all
others are varied.

\begin{figure}[ptbh]
\begin{center}
\includegraphics[width=7cm]{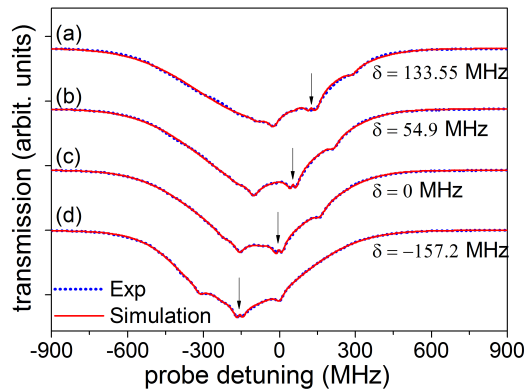}
\end{center}
\caption{(Color online) Experimental observation and simulation of the
absorption spectra by scanning the probe beam but the coupling beam locked to
different excited levels. The shift of the locking frequencies are given
relative to the transition $F=2$ to $F^{\prime}= 2$ ($\delta=0$). The coupling
laser beam power is fixed at $10$ mW. }%
\label{fig5}%
\end{figure}

We also investigate the absorption spectra with the coupling laser locked to
different SAS peaks. It is shown in Fig. \ref{fig5}. The detunings of coupling
laser are relative to the transition frequency from $F=2$ to $F^{\prime}=2$.
We can see that the VSOP dips and the EIA peak also shift over the Doppler
background with the coupling frequency shifting, and the frequency
translations are the same. It is clearly seen that as the coupling laser being
red-detuned ($\delta=-157.2$ MHz) for $F=2\rightarrow F^{\prime}=2$
transition, the magnitudes of the VSOP peaks belonging to the $V_{1}$ velocity
selective group of atoms increase, while the other VSOP dips
belonging to the $V_{2}$ velocity selective group decrease due to the
Doppler-broadening weight variation. All absorption spectra are simulated at
fitted parameters, in good agreement with the experimental spectrum, and give
the same parameter values for $\gamma_{c}$ and $\gamma_{p}$, and the
determined $\Omega_{c}$ is $17$ MHz, very close to the estimated value $17.9$
MHz from the coupling laser power $10$ mW with beam size $d=0.2$ cm. The exact
coincidence with the experimental observation indicates the reasonableness of
the theoretical model as well.

\begin{figure}[ptbh]
\begin{center}
\includegraphics[width=7cm]{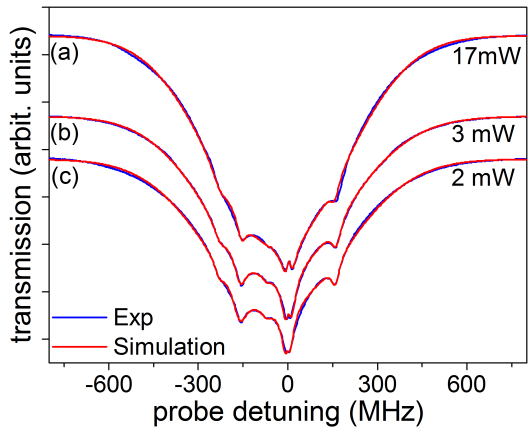}
\end{center}
\caption{(Color online) Effect of the coupling laser power on the absorption
spectra. The coupling beam is locked to the crossover transition $F=2$ to
CO$_{12}$. The probe beam power is about $0.005$ mW.}%
\label{fig6}%
\end{figure}

Finally, we studied the effect of the coupling laser pump power on the VSOP
dips and EIT peak for $^{87}$Rb atoms. The power varies from $2$ mW to $17$
mW, which are shown in Fig. \ref{fig6}. In the experiment, the coupling laser
frequency is locked to the crossover transition $F=2$ to CO$_{12}$. As we
know, the power of coupling laser only affects the EIT spectral structure,
further separating the EIT peaks, even entering into the region of
Autler-Townes splitting. The least-square fit for the spectral data also gives
the same parameter values for $\gamma_{c}$ and $\gamma_{p}$, but different
$\Omega_{c}=9$, $11.6$ and $23$ MHz at power $2$, $3$ and $17$ mW,
respectively, which are approaching to the estimated values $8$, $9.8$ and
$23$ MHz from the measured power and coupling laser beam waist size $d=0.2$ cm.

\section{conclusion}

In this paper, we reinvestigate the EIT and VSOP spectra in a five-level
$\Lambda$-scheme atomic system of $^{87}$Rb atom, where two ground hyperfine
states ($F=1,2$) of the $^{5}S_{1/2}$ level and three excited hyperfine states
($F^{\prime}=0,1,2$) corresponding to $^{5}P_{3/2}$ level are involved. Rather
than starting from the theory based on the Bloch equations, we propose a
semiempirical model to numerically explain the experimental observation. It
can give exact coincidence for the spectral profile and details.

In our model, we consider EIT and VSOPs separately where the VSOP is supposed
to have a Lorentzian-type spectral feature while EIT adopts the standard EIT
spectral structure. The effective spectrum can be obtained by the sum of these
spectra superimposed on the Doppler-broadening background. The effect of the
coupling frequency detuning and coupling beam power on the absorption spectra
is also studied. The simulated spectra show fairly good agreement with the
experimental findings. It proves the validity of our semiemperical model, and
reversely it helps us to extract useful dynamic information from the observed
spectrum. This method also can be extended to complicated cases such as the
N-, M- or Y-type systems with multi-laser fields \cite{963,1167,1216}. If the VSOP experimental configuration is adopted, principally, we can decompose the physical process as the combination of Doppler-broadening and Doppler-free ones.

\begin{acknowledgements}
This work is supported by the MOST 2013CB922003 of the National Key Basic
Research Program of China, and by NSFC (No. 91421305, No. 91121005, and No. 11174329).
\end{acknowledgements}


\end{document}